\documentstyle[epsfig,sprocl]{article}

\def\Journal#1#2#3#4{{#1} {\bf #2}, #3 (#4)}

\def\NPB{{\em Nucl. Phys.} B}
\def\PLB{{\em Phys. Lett.}  B}

\def\PR{\em Physics Reports}

\def\as{\alpha_s}
\def\LL{\Lambda}

\def\be{\begin{equation}}
\def\ee{\end{equation}}
\def\bea{\begin{eqnarray}}
\def\eea{\end{eqnarray}}

\begin{document}

\title{RENORMALONS ON THE LATTICE AND THE OPE FOR THE PLAQUETTE: 
A STATUS REPORT
\vskip-2.4cm\hfill\small UPRF-00-13\vskip2.4cm
}

\author{F. DI RENZO}

\address{Universit\'a degli Studi di Parma and 
I.N.F.N., Gruppo collegato di Parma, \\ 
viale delle Scienze, 43100 Parma, Italy \\
E-mail: direnzo@parma.infn.it} 


\maketitle\abstracts{ The first ten coefficients in the perturbative 
expansion of the plaquette in Lattice $SU(3)$ are computed both on a 
$8^4$ and on a $24^4$ lattice. They are shown to be fully consistent 
with the growth dictated by the first IR Renormalon and with the 
expected finite size effects on top of that. As already pointed out a 
few years ago, this leads to a puzzling result on the smaller lattice: 
when the contribution associated with the Renormalon is subtracted from 
Monte Carlo measurements of the plaquette, what is left over does not 
scale (as expected) as $a^4$, but as $a^2$. While the analysis is not 
yet complete on the larger lattice, the implications of such a finding 
is discussed.}

\section{Introduction: the Gluon Condensate in Lattice Gauge Theory}
A longstanding problem in Lattice Gauge Theory (LGT) is that of a 
first principles determination of the Gluon Condensate (GC). 
In LGT the GC is given in terms of Wilson Loops, for instance the basic 
plaquette $W$, for which an Operator Product Expansion (OPE) can be written 
in terms of
\begin{equation}
\label{eq:ope}
W = W_0 + \frac{W_4\,\LL^4}{Q^4} + \ldots
\end{equation}
Here $Q=1/a$ is the inverse lattice spacing, acting as the scale needed in 
every definition of the GC, $W_0$ is the contribution associated with the 
Identity operator, while $W_4$ is associated with the ``genuine'' ({\em i.e.} 
$dim=4$) condensate. In view of Eq.~(\ref{eq:ope}), a standard 
approach~\cite{pisa} was to exploit the following formula
\begin{equation}
\label{eq:piform}
W_{MC}(\beta) - \sum_n c_n \beta^{-n} = c \; Z(\beta) \; G_2 \; (a\LL)^4 
+ \ldots 
\end{equation}
which is to be understood as follows: $W_0$ is computed in Perturbation 
Theory and is subtracted from the Monte Carlo measurements at various 
values of $\beta$; the leading contribution that is left on the right 
is the second term in the OPE, which is the relevant one, whose signature 
is dictated by Asymptotic Scaling, that is $(a\LL)^4 \sim 
\exp(-\beta/(2b_0))$. \\
The former procedure is actually ill defined, since the definition of the 
perturbative contribution to be subtracted is plagued by Renormalons. In 
Ref.~\cite{8L} the first eight coefficients of the perturbative expansion 
of the plaquette were computed via Numerical Stochastic Perturbation 
Theory, the Renormalon factorial growth was singled out and shown to 
introduce an indetermination of order $(\LL/Q)^4$, which is just the order 
of the term one would be interested in. The situation is in a 
sense even worse. In Ref.~\cite{L2} the Renormalon contribution was resummed 
and subtracted: to our astonishment what was left over did not 
scale as $a^4$, but as $a^2$, at least on the $8^4$ lattice on which the 
computations were first performed on. Such a result is a 
challenge to our understanding of power effects: for a lucid discussion see 
Ref.~\cite{beneke}. 

\section{Renormalons on the lattice: new results and confirmation}
Ne now address the following questions: Was the 
leading behaviour correctly singled out? Are finite size effects under 
control (remember that the Renormalon 
growth impinges more and more on the IR region as higher order orders are 
computed)? Answers can be got (for details see Ref.~\cite{10L}) by 
considering the following formula for the Renormalon contribution 
\begin{equation}
W_0^{\rm ren}(s,N) = C \; \int^{Q^2}_{Q_0^2(N)} \;
\frac{k^2\,dk^2}{Q^4} \; \as(s k^2) = \sum_n \, C_n^{\rm ren}(C,s,N) \, 
\beta^{-n}
\end{equation}
The integral is simply the expected form (obtained from dimensional and 
Ren.G. considerations) for the condensate on a finite 
lattice: the lower limit of integration is the IR 
cutoff, function of the lattice size $N$, while the scale $s$ is 
in charge of matching from a continuum to a lattice scheme. The coefficients 
in the expansion (computable in terms of Incomplete Gamma functions) are 
functions of an overall constant, the scale $s$ and the lattice size $N$. 
In Ref.~\cite{8L} the values for $s$ and $C$ were obtained by fitting 
$C_n^{\rm ren}(C,s,N)$ to the first eight coefficients on the $N=8$ lattice. 
From $C_n^{\rm ren}(C,s,N)$ one can now infer higher 
orders on different lattice sizes (simply changing the parameter $N$). 
By doing this and actually computing the first ten coefficients of the 
expansion both on a $8^4$ and on a $24^4$ lattice, one gets for instance 
Fig~(1). While the actual numbers will be published soon 
elsewhere, an impressive agreement is manifest between the first 
ten coefficients as inferred as above and as actually computed on the 
$24^4$ lattice. Finite size effects are of order $3-4\%$ at tenth order. 

\begin{figure}[t]
\label{fig:ok}
\centerline{\psfig{figure=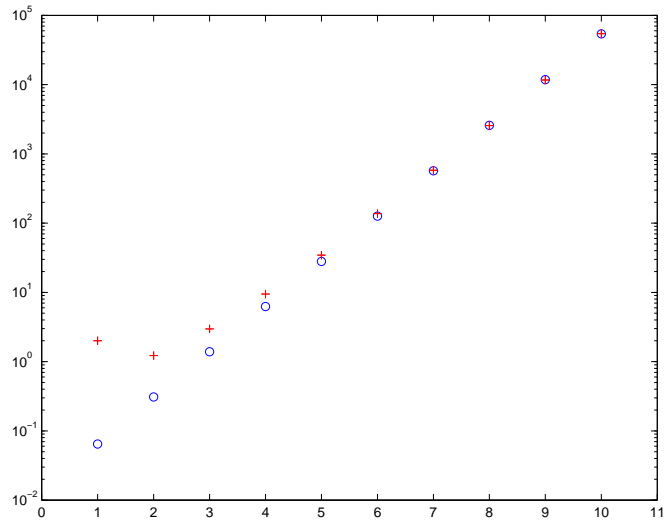,height=7cm}}
\caption{The first ten coefficients in the expansion of the plaquette 
on a $24^4$ lattice (circles) versus the expected asymptotic behaviour 
(crosses) as explained in the text.}
\end{figure}

\section{Conclusions and perspectives}
What about the subtraction? By repeating the 
procedure the bizarre result is actually stable on the $8^4$ lattice, 
which leaves the puzzle still there. On the larger lattice work is still in 
progress: first indications are that there {\em could} be space for recovering 
the standard ($a^4$) result, due to a tiny interplay between perturbative 
and non--perturbative finite size effects. This would of course still 
pose the question of what to blame for the dependence on finite size. 
Further study on the sensitivity to boundary conditions is also in progress. 

\section*{Acknowledgments}
The author is indebted to the other people involved in the NSPT project: 
G. Burgio, G. Marchesini, E. Onofri, M. Pepe, L. Scorzato.


\section*{References}
\begin{thebibliography}{99}
\bibitem{pisa}See for example B. All\'es, M. Campostrini, A. Feo and 
H. Panagopoulos, \Journal{\PLB}{324}{443}{1994} and 
references therein.
\bibitem{8L}F. Di Renzo, G. Marchesini and E. Onofri, 
\Journal{\NPB}{457}{202}{1995}.
\bibitem{L2}G. Burgio, F. Di Renzo, G. Marchesini and E. Onofri, 
\Journal{\PLB}{422}{98}{1998}.
\bibitem{beneke}M. Beneke, \Journal{\PR}{317}{1}{1999}.
\bibitem{10L}F. Di Renzo and L. Scorzato, to be issued soon. 
\end{thebibliography}
\end{document}